\documentclass{article}

\usepackage{amsmath}
\usepackage{amsfonts}
\usepackage{amssymb}
\usepackage{amscd}
\usepackage{amsthm}
\usepackage{latexsym}
\usepackage{times}
\usepackage{mathrsfs}
\usepackage{epsfig}
\usepackage{color}

\usepackage{xy}

\newcommand{\ie}{\emph{i.e.}~}

\newcommand{\cf}{\emph{cf.}~}

\def\B{\mathscr B}
\def\BB{\mathcal B}
\def\C{\mathbb C}

\def\EE{\mathcal E}
\def\F{\mathscr F}

\def\G{\mathcal G}
\def\H{\mathcal H}
\def\K{\mathcal K}

\def\M{\mathsf M}
\def\N{\mathbb N}

\def\R{\mathbb R}

\def\Z{\mathbb Z}

\def\<{\left\langle}
\def\>{\right\rangle}
\def\({\left(}
\def\){\right)}
\def\[{\left[}
\def\]{\right]}
\def\ltwo{\mathsf{L}^{\:\!\!2}}
\def\lone{\mathsf{L}^{\:\!\!1}}

\def\e{\mathop{\mathrm{e}}\nolimits}
\def\d{\mathrm{d}}

\def\Aut{\mathop{\mathrm{Aut}}\nolimits}
\def\supp{\mathop{\mathrm{supp}}\nolimits}

\def\diag{\mathop{\mathrm{diag}}\nolimits}

\def\Hom{\mathrm{Hom}}

\def\SU{\textsc{SU}}

\newtheorem{Theorem}{Theorem}[section]

\newtheorem{Lemma}[Theorem]{Lemma}
\newtheorem{Corollary}[Theorem]{Corollary}
\newtheorem{Proposition}[Theorem]{Proposition}
\newtheorem{Definition}[Theorem]{Definition}
\newtheorem{Example}[Theorem]{Example}

\begin{document}

\title{\Large\textbf{The method of the weakly conjugate operator: \\
Extensions and applications to operators on graphs and groups}}

\author{M. M\u antoiu$^1$, S. Richard$^2\footnote{On leave from Universit\'e de Lyon;
Universit\'e Lyon 1; INSA de Lyon, F-69621; Ecole Centrale de
Lyon; CNRS, UMR5208, Institut Camille Jordan, 43 blvd du 11
novembre 1918, F-69622 Villeurbanne-Cedex, France.}\ $ and R.
Tiedra de Aldecoa$^3$}
\date{\small}
\maketitle \vspace{-1cm}

\begin{quote}
\emph{
\begin{itemize}
\item[$^1$] Departamento de Matem\'aticas, Universidad de Chile, Las Palmeras 3425, Casilla 653,
Santiago, Chile
\item[$^2$] Department of Pure Mathematics and Mathematical Statistics,
Centre for Mathematical Sciences, University of Cambridge,
Cambridge, CB3 0WB, United Kingdom
\item[$^3$] CNRS (UMR 8088) and Department of Mathematics, University of Cergy-Pontoise,
2 avenue Adolphe Chauvin, 95302 Cergy-Pontoise Cedex, France
\item[] \emph{E-mails:} Marius.Mantoiu@imar.ro, sr510@cam.ac.uk,
rafael.tiedra@u-cergy.fr
\end{itemize}
  }
\end{quote}

\begin{abstract}
In this review we present some recent extensions of the method of
the weakly conjugate operator. We illustrate these developments
through examples of operators on graphs and groups.
\end{abstract}

%-------------------------------------------------------------------------------------
\section{Introduction}
\setcounter{equation}{0}
%-------------------------------------------------------------------------------------

In spectral analysis, one of the most powerful tools is the
method of the conjugate operator, also called Mourre's commutator
method after the seminal work of Mourre in the early eighties.
This approach has reached a very high degree of precision and
abstraction in \cite{ABG}; see also \cite{GGM} for further
developments. In order to study the nature of the spectrum of a
selfadjoint operator $H$, the main idea of the standard method is
to find an auxiliary selfadjoint operator $A$ such that the
commutator $i[H,A]$ is strictly positive when localized in some
interval of the spectrum of $H$. More precisely, one looks for
intervals $J$ of $\R$ such that
\begin{equation}\label{Mou}
E(J)i[H,A]E(J)\geq a E(J)
\end{equation}
for some strictly positive constant $a$ that depends on $J$, where
$E(J)$ denotes the spectral projection of $H$ on the interval $J$.
An additional compact contribution to (\ref{Mou}) is allowed, greatly
enlarging the range of applications.

When strict positivity is not available, one can instead look for
an $A$ such that the commutator is positive and injective, \ie
\begin{equation}\label{mar}
i[H,A]>0.
\end{equation}
This requirement is close to the one of the Kato-Putnam theorem,
\cf \cite[Thm. XIII.28]{RS}. A new commutator method based on such
an inequality was proposed in \cite{BKM,Boutet/Mantoiu}. By
analogy to the method of the conjugate operator, it has been
called the method of the weakly conjugate operator (MWCO). Under
some technical assumptions, both approaches lead to a limiting
absorption principle, that is, a control of the resolvent of $H$
near the real axis. In the case of the usual method of the
conjugate operator, this result is obtained locally in $J$, and
away from thresholds. The MWCO establishes the existence of the
boundary value of resolvent also at thresholds, but originally
applies only to situations where the operator $H$ has a purely
absolutely continuous spectrum. This drawback limits drastically
the range of applications. However, in some recent works
\cite{MRT07,MT07,Richard06} the MWCO has also been applied
successfully to examples with point spectrum. This review intend
to present and illustrate some of these extensions through
applications to the spectral theory of operators acting on groups
and graphs.

Compared to the huge number of applications based on an inequality
of the form \eqref{Mou}, the number of papers that contain
applications of the MWCO is very small. Let us cite for example
the works \cite{Iftimovici/Mantoiu,Mantoiu/Pascu,Mantoiu/Richard}
that deal with the original form of the theory, and the papers
\cite{FS,H} that contain very close arguments. The derivation of
the limiting absorption principle of \cite{FS} has been abstracted
in \cite{Richard06}. The framework of \cite{Richard06} is still
the one of the MWCO, but since its result applies a certain class
of two-body Schr\"odinger operators which have bound states below
zero, it can also be considered as the first extension of the MWCO
dealing with operators that are not purely absolutely continuous.
The main idea of \cite{Richard06} is that $H$ itself can add some
positivity to \eqref{mar}. The new requirement is the existence of
a constant $c\geq0$ such that $$ -cH+i[H,A]>0. $$ This inequality,
together with some technical assumptions, lead to a limiting
absorption principle which is either uniform on $\R$ if $c=0$ or
uniform on $[0,\infty)$ if $c>0$.

The extensions of the MWCO developed in \cite{MRT07,MT07} are of a
different nature. In these papers, the operators $H$ under
consideration admit a natural conjugate operator $A$ that fulfills
the inequality $i[H,A]\geq0$, namely, the commutator is positive
but injectivity may fail. In that situation, the authors
considered a decomposition $\H:=\K\oplus\G$, and the restrictions
of $H$ and $i[H,A]$ to these subspaces. In favorable circumstances
the injectivity can be restored in one of the subspace, and a
comprehensible description of the vectors of the second subspace
can be given. This decomposition leads again to statements that
are close to the ones of the MWCO, but which apply to operators
with arbitrary spectrum. This extension is described in Section
\ref{ext}.

We would also like to mention the references \cite{Bu1,Bu2}. They pereform what can be considered as a unitary
version of the MWCO and extend the Kato-Putnam analysis of unitary
operators to the case of unbounded conjugate operators. The main
applications concern time-depending propagators.

The content of this review paper is the following. In Section
\ref{mofc} we recall the original method of the weakly conjugate
operator, and then present an abstract version of the approach used
in \cite{MRT07,MT07}. In Section \ref{secad} we present
applications of this approach to the study of adjacency operators
on graphs. A similar analysis for operators
of convolution on locally compact groups is performed in Section \ref{seclcg}.

Let us finally fix some notations. Given a selfadjoint operator
$H$ in a Hilbert space $\H$, we write $\H_{\rm c}(H)$, $\H_{\rm
ac}(H)$, $\H_{\rm sc}(H)$, $\H_{\rm s}(H)$ and $\H_{\rm p}(H)$
respectively for the continuous, absolutely continuous, singularly
continuous, singular and pure point subspaces of $\H$ with respect
to $H$. The corresponding parts of the spectrum of $H$ are denoted
by $\sigma_{\rm c}(H)$, $\sigma_{\rm ac}(H)$, $\sigma_{\rm
sc}(H)$, $\sigma_{\rm s}(H)$ and $\sigma_{\rm p}(H)$.

\medskip

{\bf Acknowledgements.} M. M. was partially supported by the
Chilean Science Foundation {\it Fondecyt} under the Grant 1085162.
S. R. and R. T. d A. thank the Swiss National Science Foundation
for financial support.

%-------------------------------------------------------------------------------------
\section{The method of the weakly conjugate operator}\label{mofc}
\setcounter{equation}{0}
%-------------------------------------------------------------------------------------

In this section we recall the basic characteristics of the method
of the weakly conjugate operator, as originally introduced and applied to
partial differential operators in \cite{BKM,Boutet/Mantoiu}. We
then present the abstract form of the extension developed in
\cite{MRT07,MT07}. The method works for unbounded operators, but
for our purposes it is enough to assume $H$ bounded.

%-------------------------------------------------------------------------------------
\subsection{The standard theory}
%-------------------------------------------------------------------------------------

We start by introducing some notations. The symbol $\H$ stands for
a Hilbert space with scalar product
$\<\:\!\cdot\:\!,\:\!\cdot\:\!\>$ and norm $\|\cdot\|$. Given two
Hilbert spaces $\H_1$ and $\H_2$, we denote by $\B(\H_1,\H_2)$ the
set of bounded operators from $\H_1$ to $\H_2$, and put
$\B(\H):=\B(\H,\H)$.  We assume that $\H$ is endowed with a
strongly continuous unitary group $\{W_t\}_{t\in\R}$. Its
selfadjoint generator is denoted by $A$ and has domain $D(A)$. In
most of the applications $A$ is unbounded.

\begin{Definition}\label{defC1}
A bounded selfadjoint operator $H$ in $\H$ belongs to $C^1(A;\H)$
if one of the following equivalent condition is satisfied:
\begin{enumerate}
\item[(i)] the map $\R\ni t\mapsto W_{-t}HW_t\in\B(\H)$ is strongly differentiable,
\item[(ii)] the sesquilinear form
$$
D(A)\times D(A)\ni(f,g)\mapsto i\<Hf,Ag\>-i\<Af,Hg\>\in\C
$$
is continuous when $D(A)$ is endowed with the topology of $\H$.
\end{enumerate}
\end{Definition}

We denote by $B$ the strong derivative in (i), or equivalently the
bounded selfadjoint operator associated with the extension of the
form in (ii). The operator $B$ provides a rigorous meaning  to the
commutator $i[H,A]$. We write $B>0$ if $B$ is positive and
injective, namely if $\<f,Bf\>>0$ for all $f\in\H\setminus\{0\}$.

\begin{Definition}
The operator $A$ is \emph{weakly conjugate to} the bounded
selfadjoint operator $H$ if $H\in C^1(A;\H)$ and $B\equiv
i[H,A]>0$.
\end{Definition}

For $B>0$ let us consider the completion $\BB$ of $\H$ with
respect to the norm $\|f\|_\BB:=\<f,Bf\>^{1/2}$. The adjoint space
$\BB^*$ of $\BB$ can be identified with the completion of $B\H$
with respect to the norm $\|g\|_{\BB^*}:=\<g,B^{-1}g\>^{1/2}$. One
has then the continuous dense embeddings
$\BB^*\hookrightarrow\H\hookrightarrow\BB$, and $B$ extends to an
isometric operator from $\BB$ to $\BB^*$. Due to these embeddings
it makes sense to assume that $\{W_t\}_{t\in \R}$ restricts to a
$C_0$-group in $\BB^*$, or equivalently that it extends to a
$C_0$-group in $\BB$. Under this assumption (tacitly assumed in
the sequel) we keep the same notation for these $C_0$-groups. The
domain of the generator of the $C_0$-group in $\BB$ (resp.
$\BB^*$) endowed with the graph norm is denoted by $D(A,\BB)$
(resp.~$D(A,\BB^*)$). In analogy with Definition \ref{defC1} the
requirement $B\in C^1(A;\BB,\BB^*)$ means that the map $\R\ni
t\mapsto W_{-t}BW_t\in\B(\BB,\BB^*)$ is strongly differentiable,
or equivalently that the sesquilinear form $$ D(A,\BB)\times
D(A,\BB)\ni(f,g)\mapsto i\<f,BAg\>-i\<Af,Bg\>\in\C $$ is
continuous when $D(A,\BB)$ is endowed with the topology of $\BB$.
Here, $\<\:\!\cdot\:\!,\:\!\cdot\:\!\>$ denotes the duality
between $\BB$ and $\BB^*$. Finally let $\EE$ be the Banach space
$\big(D(A,\BB^*), \BB^*\big)_{1/2,1}$ defined by real
interpolation, see for example \cite[Proposition 2.7.3]{ABG}. One
has then the natural continuous embeddings
$\B(\H)\subset\B(\BB^*,\BB)\subset\B(\EE,\EE^*)$ and the following
results \cite[Theorem 2.1]{Boutet/Mantoiu}:

\begin{Theorem}\label{thmabstract}
Assume that $A$ is weakly conjugate to $H$ and that $B\equiv
i[H,A]$ belongs to $C^1(A;\BB,\BB^*)$. Then there exists a
constant $\textsc c>0$ such that
$$
\left|\<f,(H-\lambda\mp i\mu)^{-1}f\>\right|\leq\textsc c\|f\|_\EE^2
$$
for all $\lambda\in\R$, $\mu>0$ and $f\in\EE$. In particular the
spectrum of $H$ is purely absolutely continuous.
\end{Theorem}

The global limiting absorption principle plays an important role
for partial differential operators, \cf
\cite{BKM,Iftimovici/Mantoiu,Mantoiu/Pascu,Richard06}. However,
for the examples included sections 3 and 4, the space $\EE$
involved has a rather obscure meaning and cannot be greatly
simplified, so we shall only state the spectral results.

%-------------------------------------------------------------------------------------
\subsection{The extension}\label{ext}
%-------------------------------------------------------------------------------------

The extension proposed in \cite{MRT07,MT07} relies on the
following observation. Assume that $H$ is a bounded
selfadjoint operator in a Hilbert space $\H$. Assume also
that there exists a selfadjoint operator $A$ such that $H \in
C^1(A;\H)$, and that
\begin{equation}\label{meu}
B\equiv i[H,A]=K^2
\end{equation}
for some selfadjoint operator $K$, which is bounded by hypothesis.
It follows that $i[H,A]\geq0$ but injectivity may not be
satisfied. So let us introduce the decomposition of the Hilbert
space $\H:=\K \oplus \G$ with $\K:=\ker(K)$. The operator $B$ is
reduced by this decomposition and its restriction $B_0$ to $\G$
satisfies $B_0>0$. Formally, the positivity and injectivity of $B_0$ are
rather promising, but are obviously not sufficient for a direct
application of the MWCO to $B_0$.

However, let us already observe that some informations on $\H_{\rm
p}(H)$ can be inferred from \eqref{meu}. Indeed, it follows from
the Virial Theorem \cite[Prop. 7.2.10]{ABG} that any eigenvector $f$
of $H$ satisfies $\<f,i[H,A]f\>=0$. So
$0=\<f,K^2f\>=\|Kf\|^2$, \ie $f\in\ker(K)$, and one has
proved:

\begin{Lemma}
If $H \in C^1(A;\H)$ and $i[H,A]=K^2$, then $\H_{\rm p}(H) \subset
\K = \ker(K)$.
\end{Lemma}

Let us now come back to the analysis of $B_0$. The space $\BB$
can still be defined in analogy with
what has been presented in the previous section: $\BB$ is the
completion of $\G$ with respect
to the norm $\|f\|_\BB:=\<f,Bf\>^{1/2}$. Then, the adjoint space
$\BB^*$ of $\BB$ can be identified with the completion of $B\G$
with respect to the norm $\|g\|_{\BB^*}:=\<g,B^{-1}g\>^{1/2}$. But
in order to go further on, some compatibility has be imposed
between the decomposition of the Hilbert space and the operators
$H$ and $A$. Let us assume that both operators $H$ and $A$ are
reduced by the decomposition $\K \oplus \G$ of $\H$, and let $H_0$
and $A_0$ denote their respective restriction to $\G$. It clearly
follows from the above lemma that $H_0$ has no point spectrum.

We are now in a suitable position to rephrase Theorem
\ref{thmabstract} in our present framework. We freely use the notations
introduced above.

\begin{Theorem}\label{mo}
Assume that $H \in C^1(A;\H)$ and that $B\equiv i[H,A]=K^2$.
Assume furthermore that both operators $H$ and $A$ are reduced by
the decomposition $\ker(K) \oplus \ker(K)^\bot$ of $\H$ and that
$B_0$ belongs to $C^1(A_0;\BB,\BB^*)$. Then a limiting absorption
principle holds for $H_0$ uniformly on $\R$, and in particular the
spectrum of $H_0$ is purely absolutely continuous.
\end{Theorem}

A straightforward consequence of this statement is that
$$
\H_{\rm sc}(H) \subset \H_{\rm s}(H) \subset \K = \ker(K).
$$

We will see in the applications below that Theorem \ref{mo}
applies to various situations, and that it really is a useful
extension of the original method of the weakly conjugate operator.

%-------------------------------------------------------------------------------------
\section{Spectral analysis for adjacency operators on graphs}\label{secad}
\setcounter{equation}{0}
%-------------------------------------------------------------------------------------

We present in this section the results of \cite{MRT07} on the
spectral analysis for adjacency operators on graphs. We follow the
notations and conventions of this paper regarding graph theory.

A {\it graph} is a couple $(X,\sim)$ formed of a non-void
countable set $X$ and a symmetric relation $\sim$ on $X$ such that
$x\sim y$ implies $x\ne y$. The points $x\in X$ are called
vertices and couples $(x,y)\in X\times X$ such that $x\sim y$ are
called edges. So, for simplicity, multiple edges and loops are
forbidden in our definition of a graph. Occasionally $(X,\sim)$ is
said to be a simple graph. For any $x\in X$ we denote by
$N(x):=\{y\in X\:\!:\:\!y\sim x\}$ the set of neighbours of $x$.
We write $\deg(x):=\#N(x)$ for the degree or valence of the vertex
$x$ and $\deg(X):=\sup_{x\in X}\deg(x)$ for the degree of the
graph. We also suppose that $(X,\sim)$ is uniformly locally
finite, \ie that $\deg(X)<\infty$. A path from $x$ to $y$ is a
sequence $p=(x_0,x_1,\dots,x_n)$ of elements of $X$, usually
denoted by $x_0x_1\dots x_n$, such that $x_0=x$, $x_n=y$ and
$x_{j-1}\sim x_j$ for each $j\in\{1,\dots,n\}$.

Throughout this section we restrict ourselves to graphs $(X,\sim)$
which are simple, infinite countable and uniformly locally finite.
Given such a graph we consider the adjacency operator $H$ acting
in the Hilbert space $\H:=\ell^2(X)$ as
$$
(Hf)(x):=\sum_{y\sim x}f(y),\quad f\in\H,~x\in X.
$$
Due to \cite[Theorem 3.1]{Mohar/Woess}, $H$ is a bounded
selfadjoint operator with $\|H\|\le\deg(X)$ and spectrum
$\sigma(H)\subset[-\deg(X),\deg(X)]$.

Results on the nature of the spectrum of adjacency operators on graphs are quite
sparse. Some absolutely continuous examples are given in \cite{Mohar/Woess}, including the
lattice $\mathbb Z^n$ and homogeneous trees. For cases in which singular components are present
we refer to \cite{DS02}, \cite{GZ01}, \cite{Si} and \cite{Ve}.

We now introduce the key concept of \cite{MRT07}. Sums over the
empty set are zero by convention.

\begin{Definition}\label{adaptat}
A function $\Phi: X\to\R$ is {\rm semi-adapted to the graph} $(X,\sim)$
if
\begin{enumerate}
\item[(i)] there exists $\textsc c\ge0$ such that $|\Phi(x)-\Phi(y)|\le\textsc c\,$ for all
$x,y\in X$ with $x\sim y$,
\item[(ii)] for any $x,y\in X$ one has
\begin{equation}\label{semi}
\sum_{z\in N(x)\cap N(y)}[2\Phi(z)-\Phi(x)-\Phi(y)]=0.
\end{equation}
\end{enumerate}
If in addition for any $x,y\in X$ one has
\begin{equation}\label{full}
\sum_{z\in N(x)\cap N(y)}\[\Phi(z)-\Phi(x)\]\[\Phi(z)-\Phi(y)\]
\[2\Phi(z)-\Phi(x)-\Phi(y)\]=0,
\end{equation}
then $\Phi$ is {\rm adapted to the graph} $(X,\sim)$.
\end{Definition}

For a function $\Phi$ semi-adapted to the graph $(X,\sim)$ we
consider in $\H$ the operator $K$ given by
$$
(Kf)(x):=i\sum_{y\sim x}\[\Phi(y)-\Phi(x)\]f(y),\quad f\in\H,~x\in
X.
$$
The operator $K$ is selfadjoint and bounded due to the condition
(i) of Definition \ref{adaptat}. It commutes with $H$, as a
consequence of Condition
 (\ref{semi}). We also decompose the Hilbert space $\H$
into the direct sum $\H=\K\oplus\G$, where $\G$ is the closure of
the range $K\H$ of $K$, thus the orthogonal complement of the
closed subspace
$$
\K:=\ker(K)=\left\{f\in\H\:\!:\:\!\textstyle\sum_{y\in
N(x)}\Phi(y)f(y) =\Phi(x)\sum_{y\in N(x)}f(y)~\textrm{ for
each}~x\in X\right\}.
$$
It is shown in \cite[Sec. 4]{MRT07} that $H$ and $K$ are reduced
by this decomposition, and that their restrictions $H_0$ and $K_0$
to the Hilbert space $\G$ are bounded selfadjoint operators.

A rather straightforward application of the general theory presented in \ref{ext} gives

\begin{Theorem}[Theorem 3.2 of \cite{MRT07}]\label{unu}
Assume that $\Phi$ is a function semi-adapted to the graph
$(X,\sim)$. Then $H_0$ has no point spectrum.
\end{Theorem}

\begin{Theorem}[Theorem 3.3 of \cite{MRT07}]\label{doi}
Let $\Phi$ be a function adapted to the graph $(X,\sim)$. Then
the operator $H_0$ has a purely absolutely continuous spectrum.
\end{Theorem}

The role of the weakly conjugate operator is played by $A:=\frac{1}{2}(\Phi K+K\Psi)$
and the assumptions imposed on $\Phi$ make the general theory work.

For a certain class of admissible graphs, the result of Theorem
\ref{doi} on the restriction $H_0$ can be turned into a
statement on the original adjacency operator $H$. The notion of
admissibility requires (among other things) the graph to be
directed. Thus the family of neighbors $N(x):=\{y\in
X\:\!:\:\!y\sim x\}$ is divided into two disjoint sets $N^-(x)$
(fathers) and $N^+(x)$ (sons), $N(x)=N^-(x)\sqcup N^+(x)$. We
write $y<x$ if $y\in N^-(x)$ and $x<y$ if $y\in N^+(x)$. On
drawings, we set an arrow from $y$ to $x$ ($x\leftarrow y$) if
$x<y$, and say that the edge from $y$ to $x$ is positively
oriented.

We assume that the directed graph subjacent to $X$, from now on
denoted by $(X,<)$, is {\it admissible} with respect to these
decompositions, \ie (i) it admits a position function and (ii) it
is uniform. {\it A position function} is a function $\Phi:X\to\Z$ such
that $\Phi(y)+1=\Phi(x)$ whenever $y<x$.  It is easy to see that
it exists if and only if all paths between two points have the
same index (which is the difference between the number of
positively and negatively oriented edges). The directed graph
$(X,<)$ is called {\it uniform} if for any $x,y\in X$ the number
$\#\[N^-(x)\cap N^-(y)\]$ of common fathers of $x$ and $y$ equals
the number $\#\[N^+(x)\cap N^+(y)\]$ of common sons of $x$ and
$y$. Thus the admissibility of a directed graph is an explicit
property that can be checked directly, without making any choice.
The graph $(X,\sim)$ is admissible if there exists an admissible
directed graph subjacent to it.

\begin{Theorem}[Theorem 1.1 of \cite{MRT07}]\label{first}
The adjacency operator of an admissible graph $(X,\sim)$ is purely
absolutely continuous, except at the origin, where it may have an
eigenvalue with eigenspace
\begin{equation}\label{ma_belle_courgette}
\ker(H)=\big\{f\in\H\:\!:\:\!\textstyle\sum_{y<x}f(y)=0=\sum_{y>x}f(y)
~\textrm{ for each}~x\in X\big\}.
\end{equation}
\end{Theorem}

Many examples of periodic graphs, both admissible and
non-admissible, are presented in \cite[Sec. 6]{MRT07}. In
particular, it is explained that periodicity does not lead
automatically to absolute continuity, especially (but not only) if
the number of orbits is infinite. $D$-products of graphs, as well
as the graph associated with the one-dimensional XY Hamiltonian,
are also treated in \cite{MRT07}. We recall in Figures
\ref{graph2}, \ref{graph0}, and \ref{graph7} some two-dimensional
$\Z$-periodic examples taken from \cite[Sec. 6]{MRT07}. More
involved, $\Z^n$-periodic situations are also available.

\begin{figure}[htbp]
\begin{center}
\input{graph2.pstex_t}
\caption{\textsf{\footnotesize Example of an admissible directed
graph with $\ker(H)\neq \{0\}$}} \label{graph2} \vspace{-10pt}
\end{center}
\end{figure}

\begin{figure}[htbp]
\begin{center}
\input{graph0.pstex_t}
\caption{\textsf{\footnotesize Example of an admissible directed
graph}} \label{graph0} \vspace{-10pt}
\end{center}
\end{figure}

\begin{figure}[htbp]
\begin{center}
\input{graph4.pstex_t}
\caption{\textsf{\footnotesize Examples of admissible, directed graphs
with $\ker(H)=\{0\}$}}
\label{graph4}
\vspace{-10pt}
\end{center}
\end{figure}

\begin{figure}[htbp]
\begin{center}
\input{graph9.pstex_t}
\caption{\textsf{\footnotesize Example of an admissible, directed
graph with $\ker(H)\neq\{0\}$}} \label{graph9} \vspace{-10pt}
\end{center}
\end{figure}

\begin{figure}[htbp]
\begin{center}
\input{graph7.pstex_t}
\caption{\textsf{\footnotesize Example of a non-admissible,
adapted graph (with function $\Phi$ as indicated)}} \label{graph7}
\vspace{-10pt}
\end{center}
\end{figure}

%-----------------------------------------------------------------------------------
\section{Convolution operators on locally compact groups}\label{seclcg}
\setcounter{equation}{0}
%-----------------------------------------------------------------------------------

In this section we consider locally compact groups $X$, abelian or
not, and convolution operators $H_\mu$, acting on $\ltwo(X)$,
defined by suitable measures $\mu$ belonging to $\M(X)$, the
Banach $^*$-algebra of complex bounded Radon measures on $X$.
Using the method of the weakly conjugate operator, we determine
subspaces $\K_\mu^1$ and $\K_\mu^2$ of $\ltwo(X)$, explicitly
defined in terms of $\mu$ and the family $\Hom(X,\R)$ of
continuous group morphisms $\Phi:X\to\R$, such that $\H_{\rm
p}(H_\mu)\subset\K_\mu^1$ and $\H_{\rm s}(H_\mu)\subset\K_\mu^2$.
This result, obtained in \cite{MT07},  supplements other works on
the spectral theory of operators on groups and graphs
\cite{BG00(1),BW05,BVZ,Bre1,Bre2,DS02,Georgescu/Golenia,GZ01,HRV93(1),HRV93(2),KSS06,Kes59,M/T,Si}.

Let $X$ be a locally compact group (LCG) with identity $e$, center
$Z(X)$ and modular function $\Delta$. Let us fix a left Haar
measure $\lambda$ on $X$, using the notation $\d x:=\d\lambda(x)$.
On discrete groups the counting measure (assigning mass $1$ to
every point) is considered. The notation \emph{a.e.} stands for
``almost everywhere" and refers to the Haar measure $\lambda$.

We consider in the sequel the convolution operator $H_\mu$,
$\mu\in\M(X)$, acting in the Hilbert space
$\H:=\ltwo(X,\d\lambda)$, \ie $$ H_\mu f:=\mu\ast f, $$ where
$f\in\H$ and $$ (\mu\ast
f)(x):=\int_X\d\mu(y)\,f(y^{-1}x)\quad\hbox{for \emph{a.e.} }x\in
X. $$ The operator $H_\mu$ is bounded with norm
$\|H_\mu\|\le\|\mu\|$, and it admits an adjoint operator $H_\mu^*$
equal to $H_{\mu^*}$, the convolution operator by $\mu^*\in\M(X)$
defined by $\mu^*(E)=\overline{\mu(E^{-1})}$. If the measure $\mu$
is absolutely continuous w.r.t. the left Haar measure $\lambda$,
so that $\d\mu=a\,\d\lambda$ with $a\in\lone(X)$, then $\mu^*$ is
also absolutely continuous w.r.t. $\lambda$ and
$\d\mu^*=a^*\d\lambda$, where
$a^*(x):=\overline{a(x^{-1})}\Delta(x^{-1})$ for \emph{a.e.} $x\in
X$. In such a case we simply write $H_a$ for $H_{a\;\!\d\lambda}$.
We shall always assume that $H_\mu$ is selfadjoint, that is
$\mu=\mu^*$.

Given $\mu\in\M(X)$, let $\varphi:X\to\R$ be such that the linear
functional $$ F:C_0(X)\to\C,\quad
g\mapsto\int_X\d\mu(x)\,\varphi(x)g(x) $$ is bounded. Then there
exists a unique measure in $M(X)$ associated to $F$, due to the
Riesz-Markov representation theorem. We write $\varphi\mu$ for
this measure, and we simply say that $\varphi$ is such that
$\varphi\mu\in\M(X)$. We call \emph{real character} any continuous
group morphism $\Phi:X\to\R$.

\begin{Definition}\label{admis}
Let $\mu=\mu^*\in\M(X)$.
\begin{enumerate}
\item[(a)] A real character $\Phi$ is \emph{semi-adapted to $\,\mu$}
if $\Phi\mu,\Phi^2\mu\in\M(X)$, and
$(\Phi\mu)\ast\mu=\mu\ast(\Phi\mu)$. The set of real characters
that are semi-adapted to $\mu$ is denoted by $\Hom^1_\mu(X,\R)$.
\item[(b)] A real character $\Phi$ is \emph{adapted to $\mu$} if $\Phi$ is semi-adapted
to $\mu,\Phi^3\mu\in\M(X)$, and
$(\Phi\mu)\ast(\Phi^2\mu)=(\Phi^2\mu)\ast(\Phi\mu)$. The set of
real characters that are adapted to $\mu$ is denoted by
$\Hom^2_\mu(X,\R)$.
\end{enumerate}
\end{Definition}

Let $\K^j_\mu:=\bigcap_{\Phi\in\Hom^j_\mu(X,\R)}\ker(H_{\Phi
\mu})$, for $j=1,2$; then the main result is the following.

\begin{Theorem}[Theorem 2.2 of \cite{MT07}]\label{princ}
Let $X$ be a LCG and let $\mu=\mu^*\in\M(X)$. Then $$ \H_{\rm
p}(H_\mu)\subset\K^1_\mu\qquad{\rm and}\qquad\H_{\rm
s}(H_\mu)\subset\K^2_\mu. $$
\end{Theorem}

The cases $\K_\mu^1=\{0\}$ or $\K_\mu^2=\{0\}$ are interesting; in
the first case $H_\mu$ has no eigenvalues, and in the second case
$H_\mu$ is purely absolutely continuous. A more precise result is
obtained in a particular situation.

\begin{Corollary}[Corollary 2.3 of \cite{MT07}]\label{precis}
Let $X$ be a LCG and let $\mu=\mu^*\in\M(X)$. Assume that there
exists a real character $\Phi$ adapted to $\mu$ such that $\Phi^2$
is equal to a nonzero constant on $\supp(\mu)$. Then $H_\mu$ has a
purely absolutely continuous spectrum, with the possible exception
of an eigenvalue located at the origin, with eigenspace
$\ker(H_\mu)=\ker(H_{\Phi\mu})$.
\end{Corollary}
Corollary \ref{precis} specially applies to adjacency operators on
certain classes of Cayley graphs, which are Hecke-type operators
in the regular representation, thus convolution operators on
discrete groups.

To see how the method of the weakly conjugate operator comes into
play, let us sketch the proof of the second inclusion in Theorem
\ref{princ}. In quantum mechanics in $\R^d$, the position
operators $Q_j$ and the momentum operators $P_j$ satisfy the
relation $P_j=i[H,Q_j]$ with $H:=-\frac12\Delta$, and the usual
conjugate operator is the generator of dilations
$D:=\frac12\sum_j(Q_jP_j+P_jQ_j)$. So if we regard
$\Phi\in\Hom^2_\mu(X,\R)$ as a position operator on $X$, it is
reasonable to think of $K:=i[H_\mu,\Phi]$ as the corresponding
momentum operator and to use $A:=\frac12(\Phi K+K\Phi)$ as a
tentative conjugate operator. In fact, simple calculations using
the hypotheses of Definition \ref{admis}.(b) show that $$
K=-iH_{\Phi\mu}\in\B(\H)\qquad{\rm and}\qquad i[H_\mu,A]=K^2. $$
Therefore $H_\mu\in C^1(A;\H)$ and the commutator $i[H_\mu,A]$ is
a positive operator. However, in order to apply the theory of
Section \ref{mofc}, we need strict positivity. So, we consider the
subspace $\G:=[\ker(K)]^\perp$ of $\H$, where $i[H_\mu,A]\equiv
K^2$ is strictly positive. The orthogonal decomposition
$\H:=\K\oplus\G$ with $\K:=\ker(K)$, reduces $H_\mu$, $K$, and
$A$, and their restrictions $H_0$, $K_0$, and $A_0$ to $\G$ are
selfadjoint. It turns out that all the other conditions necessary
to apply Theorem \ref{mo} can also be verified in the
Hilbert space $\G$. So we get the inclusion $$ \H_{\rm
s}\subset\K=\ker(H_{\Phi\mu}). $$ Since $\Phi\in\Hom^2_\mu(X,\R)$
is arbitrary, this implies the second inclusion of Theorem
\ref{princ}.

%-----------------------------------------------------------------------------------
\subsection{Examples}\label{Examples}
\setcounter{equation}{0}
%-----------------------------------------------------------------------------------

The construction of weakly conjugate operators for $H_\mu$ relies on real characters.
So a small vector space $\Hom(X,\mathbb R)$ is an obstacle to applying the method.
A real character $\Phi$ maps compact subgroups of $X$ to the unique compact subgroup
$\{0\}$ of $\mathbb R$. Consequently, abundancy of compact elements (elements $x\in X$
generating compact subgroups) prevents us from constructing weakly conjugate operators.
The extreme case is when $X$ is itself compact, so that $\Hom(X,\mathbb R)=\{0\}$.
Actually in such a case all convolution operators $H_\mu$ have pure point spectrum.
We review now briefly
some of the groups for which we succeeded in \cite{MT07} to apply the MWCO.

If the group $X$ is unimodular, one can exploit commutativity in a
non-commutative setting by using central measures (\ie elements of
the center $Z[\M(X)]$ of the convolution Banach $^*$-algebra
$\M(X)$). For instance, in the case of central groups \cite{GM67},
we have the following result ($\B(X)$ stands for the closed
subgroup generated by the set of compact elements of $X$):

\begin{Proposition}[Proposition 4.2 of \cite{MT07}]\label{centreaza}
Let $X$ be a central group and $\mu_0=\mu_0^*\in\M(X)$ a central
measure such that $\supp(\mu_0)$ is compact and not included in
$\B(X)$. Let $\mu_1=\mu_1^*\in\M(X)$ with
$\supp(\mu_1)\subset\B(X)$ and set $\mu:=\mu_0+\mu_1$. Then
$\H_{\rm ac}(H_\mu)\ne\{0\}$.
\end{Proposition}

Let us recall three examples deduced from Proposition
\ref{centreaza} taken from \cite{MT07}.

\begin{Example}
Let $X:=S_3\times\Z$, where $S_3$ is the symmetric group of degree
$3$. The group $S_3$ has a presentation $\<a,b\mid
a^2,b^2,(ab)^3\>$, and its conjugacy classes are
$E_1=E_1^{-1}=\{e\}$, $E_2=E_2^{-1}=\{a,b,aba\}$ and
$E_3=E_3^{-1}=\{ab,ba\}$. Set $\mathcal E:=\{E_2,E_3\}$ and choose
$I_{E_1},I_{E_2}$ two finite symmetric subsets of $\Z$, each of
them containing at least two elements. Then $\H_{\rm
ac}(H_{\chi_S})\ne\{0\}$ if $S:=\bigcup_{E\in\mathcal E}E\times
I_E$.
\end{Example}

\begin{Example}
Let $X:=\SU(2)\times\R$, where $\SU(2)$ is the group (with Haar
measure $\lambda_2$) of $\,2\times2$ unitary matrices of
determinant $+1$. For each $\vartheta\in[0,\pi]$ let
$C(\vartheta)$ be the conjugacy class of the matrix
$\diag(\e^{i\vartheta},\e^{-i\vartheta})$ in $\SU(2)$. A direct
calculation (using for instance Euler angles) shows that
$\lambda_2\(\bigcup_{\vartheta\in J}C(\vartheta)\)>0$ for each
$J\subset[0,\pi]$ with nonzero Lebesgue measure. Set
$E_1:=\bigcup_{\vartheta\in(0,1)}C(\vartheta)$,
$E_2:=\bigcup_{\vartheta\in(2,\pi)}C(\vartheta)$, $\mathcal
E:=\{E_1,E_2\}$, $I_{E_1}:=(-1,1)$, and
$I_{E_2}:=(-3,-2)\cup(2,3)$. Then $\H_{\rm
ac}(H_{\chi_S})\ne\{0\}$ if $S:=\bigcup_{E\in\mathcal E}E\times
I_E$.
\end{Example}

\begin{Example}
Let $X$ be a central group, let $z\in Z(X)\setminus\B(X)$, and set
$\mu:=\delta_z+\delta_{z^{-1}}+\mu_1$ for some
$\mu_1=\mu_1^*\in\M(X)$ with $\supp(\mu_1)\subset\B(X)$. Then
$\mu$ satisfies the hypotheses of Proposition \ref{centreaza}, and
we can choose $\Phi\in\Hom(X,\R)$ such that
$\Phi(z)=\frac12\Phi(z^2)\ne0$ (note in particular that
$z\notin\B(X)$ iff $z^2\notin\B(X)$ and that $\Phi\mu_1=0$). Thus
$\H_{\rm s}(H_\mu)\subset\ker(H_{\Phi\mu})$. But $f\in\H$ belongs
to
$\ker(H_{\Phi\mu})=\ker\big(H_{\Phi(\delta_z+\delta_{z^{-1}})}\big)$
iff $f(z^{-1}x)=f(zx)$ for {\em a.e.} $x\in X$. This periodicity
w.r.t. the non-compact element $z^2$ implies that the
$\ltwo$-function $f$ should vanish {\em a.e.} and thus that
$\H_{\rm ac}(H_\mu)=\H$.
\end{Example}

If $X$ is abelian all the commutation relations in Definition
\ref{admis} are satisfied. Moreover one can use the Fourier
transform $\F$ to map unitarily $H_\mu$ on the operator $M_m$ of
multiplication with $m=\F(\mu)$ on the dual group $\widehat X$ of
$X$. So one gets from from Theorem \ref{princ} a general lemma on
muliplication operators. We recall some definitions before stating
it.

\begin{Definition}
The function $m:\widehat X\to\C$ is \emph{differentiable at
$\xi\in\widehat X$ along the one-parameter subgroup
$\varphi\in\Hom(\R,\widehat X)$} if the function $\R\ni t\mapsto
m(\xi+\varphi(t))\in\C$ is differentiable at $t=0$. In such a case
we write $\(d_\varphi m\)(\xi)$ for $\frac\d{\d
t}\;\!m(\xi+\varphi(t))\big\vert_{t=0}$. Higher order derivatives,
when existing, are denoted by $d_{\varphi}^km$, $k\in\N$.
\end{Definition}

We say that the one-parameter subgroup $\varphi:\R\to\widehat X$
is in $\Hom_m^1(\R,\widehat X)$ if $m$ is twice differentiable
w.r.t. $\varphi$ and $d_\varphi m,d^2_\varphi m\in\F(\M(X))$. If,
in addition, $m$ is three times differentiable w.r.t. $\varphi$
and $d^3_\varphi m\in\F(\M(X))$ too, we say that $\varphi$ belongs
to $\Hom_m^2(\R,\widehat X)$.

\begin{Lemma}[Corollary 4.7 of \cite{MT07}]\label{babel}
Let $X$ be a locally compact abelian group and let $m_0,m_1$ be
real functions with $\F^{-1}(m_0),\F^{-1}(m_1)\in\M(X)$ and
$\supp(\F^{-1}(m_1))\subset\B(X)$. Then $$ \H_{\rm
p}(M_{m_0+m_1})\subset\bigcap_{\varphi\in\Hom_{m_0}^1(\R,\widehat
X)}\ker(M_{d_\varphi m_0}) $$ and $$ \H_{\rm
s}(M_{m_0+m_1})\subset\bigcap_{\varphi\in\Hom^2_{m_0}(\R,\widehat
X)}\ker(M_{d_\varphi m_0}). $$
\end{Lemma}

We end up this section by considering a class of semidirect
products. Let $N,G$ be two discrete groups with $G$ abelian (for
which we use additive notations), and let $\tau:G\to\Aut(N)$ be a
group morphism. Let $X:=N\times_\tau G$ be the $\tau$-semidirect
produt of $N$ by $G$. The multiplication in $X$ is defined by $$
(n,g)(m,h):=(n\tau_g(m),g+h), $$ so that $$
(n,g)^{-1}=(\tau_{-g}(n^{-1}),-g). $$ In this situation it is
shown in \cite{MT07} that many convolution operators $H_a$, with
$a:X\to\C$ of finite support, have a non-trivial absolutely
continuous component. For instance, we have the following for a
type of wreath products.

\begin{Example}
Take $G$ a discrete abelian group and put $N:=R^J$, where $R$ is
an arbitrary discrete group and $J$ is a finite set on which $G$
acts by $(g,j)\mapsto g(j)$. Then $\tau_g\big(\{r_j\}_{j\in
J}\big):=\{r_{g(j)}\}_{j\in J}$ defines an action of $G$ on $R^J$,
thus we can construct the semidirect product $R^J\times_\tau G$.
If $G_0=-G_0\subset G$ and $R_0=R_0^{-1}\subset R$ are finite
subsets with $G_0\cap[G\setminus\B(G)]\ne\varnothing$, then
$N_0:=R_0^J$ satisfies all the conditions of \cite[Sec.
4.4]{MT07}. Thus $\H_{\rm ac}(H_{\chi_S})\ne\{0\}$ if
$S:=N_0\times G_0$.
\end{Example}

Virtually the methods of \cite{MT07} could also be applied to
non-split group extensions.

%-----------------------------------------------------------------------------------
%Bibliography
%-----------------------------------------------------------------------------------

\end{document}